\documentclass[a4paper,11pt]{article}
\pdfoutput=1 

\usepackage{jcappub} 
\usepackage[T1]{fontenc}

\usepackage{natbib}
\usepackage{pifont}
\usepackage{ulem}
\usepackage{cancel}
\usepackage{enumitem}    
\usepackage{amsthm}         
\usepackage{amsmath} 	   
\usepackage{amssymb}       
\usepackage{amsfonts}
\usepackage{graphicx} 	    
\usepackage{verbatim}
\usepackage{color}
\usepackage{colortbl}
\usepackage{float}
\usepackage{multirow}

\allowdisplaybreaks

\usepackage{subcaption}

\definecolor{babyblue}{rgb}{0.54, 0.81, 0.94}
\definecolor{corn}{rgb}{0.98, 0.93, 0.36}

\title{\huge Bouncing Cosmology made simple}

\author[a]{Anna Ijjas}
\author[b,c]{and Paul J. Steinhardt}
\affiliation[a]{Center for Theoretical Physics, Columbia University, New York, NY 10027, USA}
\affiliation[b]{Department of Physics \& Princeton Center for Theoretical Science, Princeton University, Princeton, NJ 08544, USA}
\affiliation[c]{Center for Cosmology and Particle Physics, Department of Physics,
New York University, New York, NY, 10003, USA }

\emailAdd{anna.ijjas@columbia.edu}
\emailAdd{steinh@princeton.edu}

\abstract{
We introduce  the ``wedge diagram,'' an intuitive way to illustrate how cosmological models with a classical (non-singular) bounce generically resolve fundamental problems in cosmology.  These include the well-known horizon, flatness, and inhomogeneity problems; the small tensor-to-scalar ratio observed in the cosmic microwave background; the low entropy at the beginning of a hot, expanding phase; and the avoidance of quantum runaway.  The same diagrammatic approach can be used to compare with other cosmological scenarios.
\newline\\ \today
}

\begin{document}
\maketitle 
\flushbottom


\section{Introduction}

In recent years, there has been increasing interest in cosmological models that replace the cosmological singularity (or ``big bang'') with a  ``big bounce'' -- a smooth transition from contraction to expansion -- in order to resolve fundamental problems in cosmology.   Some earlier  explorations  of this idea invoked elements like branes and extra dimensions \cite{Khoury:2001wf,Khoury:2001bz,Turok:2004gb,McFadden:2005mq,Lehners:2007nb,Graham:2017hfr}; string gas \cite{Brandenberger:1988aj,Nayeri:2005ck};  unstable field trajectories \cite{Lehners:2007ac,Tolley:2007nq}; generation of significant non-gaussianity \cite{Buchbinder:2007at,Lehners:2009qu}; or finely-tuned  initial conditions during the contraction phase \cite{Gasperini:1992em,Brandenberger:2012zb,Levy:2016xcl}; etc.  See Ref.~\cite{Brandenberger:2016vhg} for a review of these earlier scenarios.

We have since learned that many of these features are not necessary and/or only occur in particular constructions or specially fine-tuned examples \cite{Li:2013hga,Fertig:2013kwa,Ijjas:2013sua,Ijjas:2014fja,Levy:2015awa}.  Nevertheless, the earlier examples have been instructive in pointing the way to the arguably simpler approach to bouncing cosmology that we will consider here.  These  theories are based on more ordinary elements -- three spatial dimensions, scalar fields and, most importantly, a {\it non-singular} bounce that occurs at densities well below the Planck scale where quantum gravity effects are small.   In fact, not only the bounce, but the entire background evolution of these geodesically complete cosmological scenarios is described to leading order by classical equations of motion.  

The goal of this paper is to consider the generic properties and advantages of these classical types of bouncing cosmologies.  For this purpose, we utilize a hydrodynamic description \cite{Wang:1997cw,Gratton:2003pe,Khoury:2003vb,Ijjas:2013sua} of the stress-energy that is relatively insensitive to the details of the microphysics and introduce a visualization method (the ``wedge diagram'') to describe the cosmic evolution.  With these tools, we explain how classical (non-singular) bouncing cosmologies generically:

\begin{itemize}
\renewcommand{\labelitemi}{--}
\item avoid the cosmic singularity problem;
\item enable geodesically complete evolution;
\item eliminate chaotic mixmaster behavior; 
\item resolve the horizon problem;
\item explain the smoothness and flatness of the universe;
\item generate superhorizon-scale density fluctuations;
\item suppress the amplitude of (primary) tensor fluctuations;
\item avoid quantum runaway (a.k.a. the multiverse problem);
\item and naturally explain the small entropy at the onset of the expanding phase.
\end{itemize}

The basic cosmological principles underlying the wedge diagram are introduced in Section 2.  As a first example, we diagram the standard expanding big bang model in Section 3.  Section 4 constructs the wedge diagram for a non-singular bouncing cosmology.   In Section 5, we use the diagram to illustrate each of the properties described above.  For the purposes of comparison, the Appendix presents the analogous representation of big bang inflationary models.

\section{Patch size, horizon size, and cosmic curvature}

Observations show that the universe has been expanding and cooling for about 13.8 Gyrs and that the complex structure observed today emerged from a condition that was remarkably uniform when the universe was about a thousandth of its current radius, as demonstrated by cosmic microwave background (CMB) maps of the last scattering surface \cite{Komatsu:2008hk,Ade:2013lta,Sievers:2013ica}.  The only deviation from perfect uniformity was a nearly scale-invariant spectrum of density fluctuations with rms amplitude  ${\cal O}(10^{-5})$.  
 A condition for any theory to be viable is that it  agree with these observed conditions; but theories disagree on when and how the large-scale smoothness and the small-amplitude inhomogeneities arose and what will happen in the future.

To visualize the problem and different theoretical possibilities, we will track two quantities that characterize a smooth and isotropic universe in general relativity: the scale factor $a(t)$ and the Hubble parameter $H(t)$, both functions of time. 

The scale factor is a dimensionless quantity that describes how much a patch of space changes in size due to expansion (or contraction).  

\begin{quote}
{\it If the observable universe corresponds to a patch of space with radius $R(t_0)$ today ($t=t_0$), then the \underline{patch size} at any other time is equal to $\big( a(t)/a(t_0) \big) \times R(t_0)$.}
\end{quote}

The Hubble parameter, $H \equiv \dot{a}/a$ (where dot denotes differentiation with respect to time $t$), measures the expansion rate.  Assuming a smooth and isotropic universe described by the Friedmann-Robertson-Walker metric with scale factor $a(t)$,
\begin{equation}
{\rm d}s^2 = -{\rm d}t^2 + a^2(t) {\rm d}x_i {\rm d}x^i,
\end{equation} 
the stress-energy must take the form of a perfect fluid fully described by an energy density $\rho$ and pressure $p$; and the  Einstein equations can be reduced to a single constraint equation, 
\begin{equation} \label{F1}
H^2 = \left(\frac{\dot{a}}{a}\right)^2 = \frac{8 \pi G_{\rm N}}{3}\rho - \frac{k}{a^2}
\,, 
\end{equation}
and a second-order equation of motion,
\begin{equation} 
\label{F2}
\frac{\ddot{a}}{a} = -\frac{4 \pi G_{\rm N}}{3} (\rho+ 3p)\,,
\end{equation}
which together constitute the Friedmann equations.
Here $G_{\rm N}$ is Newton's constant and $k/a^2$ is the spatial curvature.  To good approximation, the universe is spatially flat ($k=0$) today;  in that limit, $a(t)$ can be normalized so that it is unity today, $a(t_0)=1$, which we will assume for the remainder of this paper. (Throughout, we set the speed of light $c \equiv 1$.)

The Friedmann equations can be combined to show that 
\begin{equation} \label{rho}
\rho(a) = \rho_0 \exp \left(-2 \int \epsilon \, {\rm d} \ln a \right) \approx \frac{\rho_0}{a^{2 \epsilon}}\,,
\end{equation} 
where
\begin{equation}
\epsilon \equiv \frac{3}{2}\left(1+ \frac{p}{\rho}\right)
\end{equation}
characterizes the equation of state and $\rho_0$ is the current density.   The value of $\epsilon$ can be interpreted as a measure of the pressure: $\epsilon$ increases with the pressure of the universe.  Positive pressure corresponds to $\epsilon>3/2$ and negative pressure corresponds to $\epsilon < 3/2$.   If the universe is dominated by a cosmological constant, $\epsilon \rightarrow 0$.

The ``$\approx$'' in Eq.~\eqref{rho} becomes an equality in the limit that $\epsilon$ is  constant.  In practice, 
its value is nearly constant over long epochs, varying rapidly from one constant value to another  when the dominant form of energy changes. For example, in the standard big bang model, the universe is radiation-dominated  (corresponding to $\epsilon = 2$) for the first 50,000 yrs after the big bang and dust-dominated (corresponding to $\epsilon=3/2$) for the remaining 13.8 Gyrs.   When the variations in $\epsilon$ are so small that they make no qualitative difference to our discussion below, we will treat $\epsilon$ as constant to simplify expressions.

Let us note two important facts about pressure and energy density that often cause confusion.  First, the pressure $p$  can be negative but, according to the constraint in Eq.~\eqref{F1}, general relativity requires that the total energy density $\rho \ge 0$.  For example, violating the {\it null energy} condition does not mean negative $\rho$; rather, it means the pressure is sufficiently negative such that $p+\rho <0$, while $\rho \ge 0$.  Second, it is generally {\it not} the case that $c_s \equiv (dp/d\rho)^{1/2}$, the sound speed  for a fluid, is equal to $p/\rho$.  
For example, if the fluid is associated with a scalar field with canonical kinetic energy density, then $c_s^2=1$ but $\epsilon \equiv \frac{3}{2} (1+p/\rho)$ can vary from zero to arbitrarily large positive values depending on its potential energy density.\footnote{More precisely, if the scalar field kinetic energy density is ${\rm KE}=\frac{1}{2} \dot{\phi}^2$ and its potential energy density is ${\rm PE}=V(\phi)$, then the pressure is given by $p= {\rm KE} - {\rm PE}$ and the total energy density is given by $\rho = {\rm KE} + {\rm PE}$; 
consequently,  $\epsilon\equiv\frac{3}{2} \left(1+ \frac{p}{\rho} \right) = 3\, \frac{{\rm KE}}{{\rm KE} + {\rm PE}}$.  The total energy density, ${\rm KE} + {\rm PE}$ must be positive, but it can be arbitrarily small; hence, $\epsilon$ can vary between zero and arbitrarily large values.}

A key concept in general relativity is the Hubble horizon, with Hubble radius or {\it horizon size} $H^{-1}$.   In an expanding big bang universe (where $\epsilon>1$), it is equal up to a factor  of order unity to the particle horizon size, 
\begin{equation}
a(t) \int_0^t \frac{{\rm d} t'}{a(t')},
\end{equation}
the maximum distance  light or particles could travel in the age of the universe and, hence, the maximum distance an observer can see at time $t$.  For the purposes of this discussion, we will ignore the proportionality factor of ${\cal O}(1)$ and treat the  particle horizon size as equal to the Hubble horizon, $H^{-1}$, during the expanding big bang phase.
  
Today, the horizon size and the patch size of the observable universe are equal, namely $R(t_0)$.  Observational evidence shows that the spatial curvature today  is negligible.  Consequently, according to the Friedmann equation, the horizon size is $H^{-1} \propto a^{\epsilon}$. 
\vspace{-.3in}
\begin{quote}
\item {\it If the horizon size is $R(t_0)$ today,  the \underline{horizon size}  at other times   is $a^{\epsilon} \times R(t_0)$.}
\end{quote}
Note that the relation between the patch size and $a(t)$ is purely geometric, essentially a definition of what it means for $a(t)$ to represent the expansion or contraction.  By contrast, the existence of the horizon and the relation between the horizon size, $a(t)$ and the equation of state $\epsilon$ is where general relativity comes into the picture. 

A third important feature is the {\it cosmic curvature factor}, 
\begin{equation}
\Omega_c \equiv -\frac{k}{a^2}\times H^{-2} \propto \frac{a^{2 \epsilon}}{a^2}\,,
\end{equation} 
a time-dependent, dimensionless quantity that characterizes the {\it apparent} spatial curvature.  Its value today is less than 0.3\% according to recent observations.
One way to think of $\Omega_c$ is as follows: The radius of curvature changes in proportion to the scale factor, $a(t)$, but the {\it apparent} curvature depends on how this distance compares to how far one can see, namely the horizon size.  The cosmic curvature factor is, up to a constant coefficient, a measure of (horizon size/radius of curvature)$^2$,  which is, in turn, proportional to a ratio we well define as  ${\cal \eta}^2 \equiv a^{2 \epsilon}/a^2$ given the two italicized relations above.\footnote{Note that $\eta$ is proportional to conformal time for constant $\epsilon$:   that is,  $d \eta \propto dt/a$ implies $\eta \propto a^{\epsilon}/a$.} 
\begin{quote}
{\it If $\Omega_c(t_0)$ is the current value of the \underline{cosmic curvature factor}, the value at other times is  $ (a^{2 \epsilon}/a^2) \times \Omega_c(t_0)$}.
\end{quote}

Analogously, the {\it cosmic anisotropy factor},  
\begin{equation} \label{anis}
\Omega_a \equiv \frac{\sigma^2}{a^6}\times H^{-2} \propto \frac{a^{2 \epsilon}}{a^6}\,,
\end{equation} 
is a time-dependent, dimensionless quantity that characterizes the {\it apparent} anisotropy.  Note the additional powers of $a(t)$ in the denominator of the anisotropy factor  $\Omega_a$ compared to the curvature factor $\Omega_c$.

\section{The cosmic wedge diagram and the expanding big bang universe}

The cosmic wedge diagram is a useful way of representing the relationship between the patch size and the horizon size.
\begin{figure}[tb]
  \centering
\includegraphics[width=.55\linewidth]{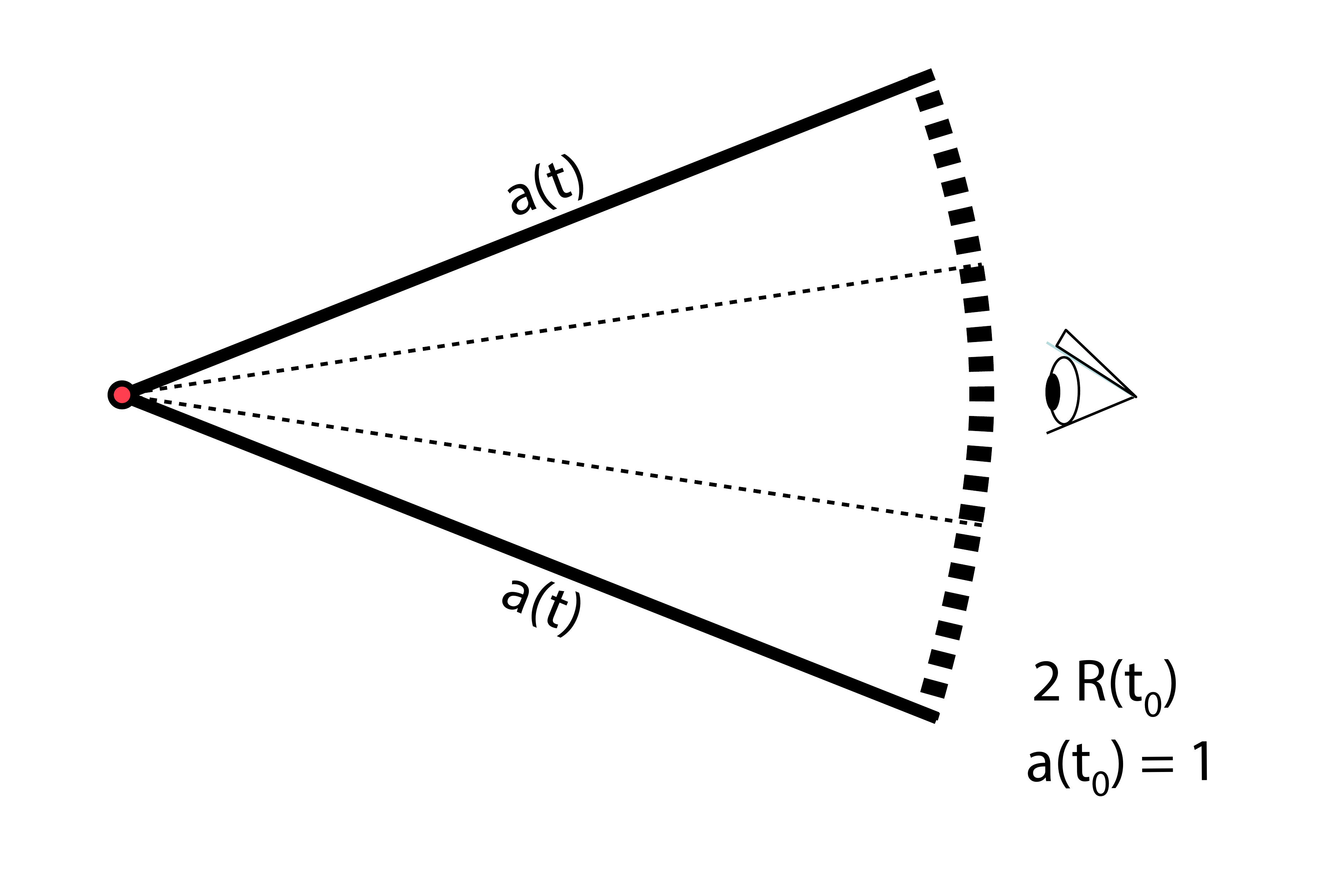}
\caption{Outline of the wedge diagram for the big bang model.  Evolution of a patch size proceeds from the big bang singularity (vertex on the left) to the present ($t=t_0$).  For example, the thick dashed curve might represent the physical size of the observable universe today, in which case arcs cut closer to the vertex correspond to the patch size at earlier times (proportional to $a(t)$).  The thin dotted lines would then represent wedge describing the evolution of a smaller patch of space.
}
\label{fig:1}
\end{figure}
The sides of the wedge (solid lines meeting at an angle in Fig.~1) are labeled by the scale factor $a(t)$.   The linear size of a patch of space at a given time $t$ is represented by an arc connecting opposites sides of the wedge. 
The wedge representation automatically encapsulates the condition that the size of a given patch of space grows linearly with $a(t)$ between the big bang (the vertex on the left where $a=0$) and today (the outermost arc where the scale factor is equal to $a(t_0)$). 

For example, we might imagine the outer (rightmost) edge of the wedge as representing the entire observable universe today, a patch with radius $R(t_0)$; in this case, with an observer at the center, the  arc length at the outer edge corresponds to the diameter with length  $2 R(t_0)$.  The same patch of space at earlier times is represented by arcs closer to the vertex.   If we want to consider at the same time the evolution of a patch smaller than the observable universe today, it would be represented by a narrower wedge, as illustrated by the dotted lines in Figure~1.  A larger slice that includes our observable universe plus space that lies beyond would be represented by a wider slice.  

All wedges meet at the same vertex, corresponding to $a=0$.  The existence of this vertex means that the cosmology is {\it geodesically incomplete}.  In the standard big bang model, the vertex corresponds to the {\it cosmic singularity}, a point in spacetime where, formally, the density and temperature diverge and the geometry shrinks to zero, a sign that the equations used to describe the evolution of the universe cannot be trusted near this point.  Finding a suitable fix is the {\it cosmic singularity problem.}

Next, we superpose on the wedge how the horizon size varies with $a(t)$; see Figure~2.  
\begin{figure}[tb]
  \centering
\includegraphics[width=.725\linewidth]{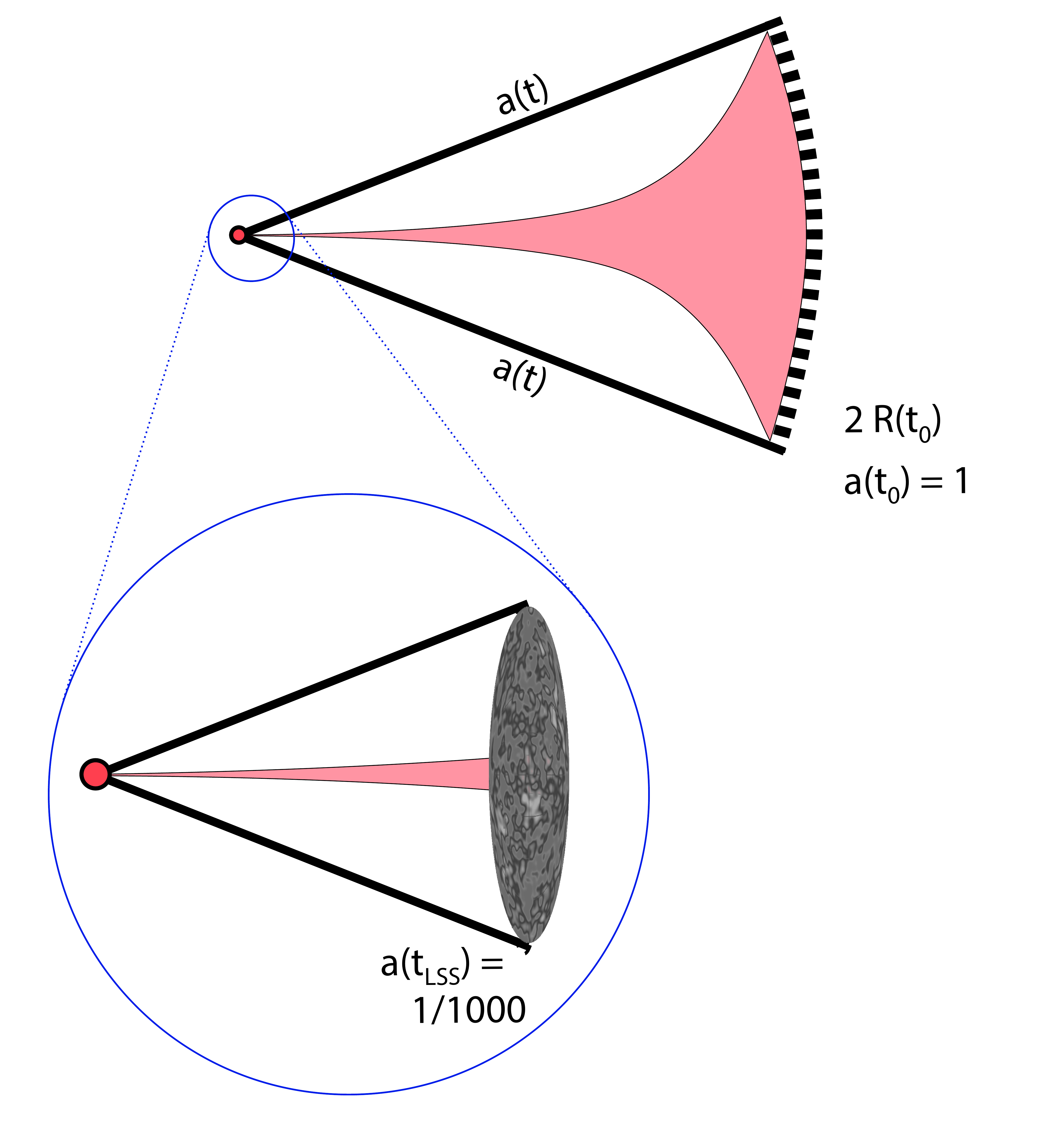}
\caption{Wedge diagram for the standard big bang model, comparing the patch size (the full wedge) to the horizon size (shaded region).  Horizon size ($\propto a^{\epsilon}$)  corresponds to the arc length across the shaded region.  Patch size ($\propto a$) corresponds to the arc length spanning the wedge edges.  The inset magnifies the corner of the wedge where $a(t)\lesssim 1/1000$, highlighting the period between the big bang and the last scattering surface (LSS).  Note that the  patch size of the last scattering surface is larger than the horizon size.
} 
\label{fig:2}
\end{figure}  
If the outer arc represents the observable universe today, then the patch size and horizon size are both equal to $R(t_0)$ at $t_0$.  At earlier times, the patch size scales as $a(t)$ and the horizon size scales as $a^{\epsilon}(t)$.  In the standard big bang model, $\epsilon>1$ for both the radiation- and dust-dominated epochs.  Consequently,  extrapolating back in time, the ratio of the horizon size to the patch size  ($\propto a^{\epsilon}/a$)  approaches zero as $a \rightarrow 0$.    Both the horizon size and patch size approach zero at the cosmological singularity, but, significantly, the horizon size approaches zero faster. 

Since the horizon represents the maximal region that is ``causally connected'' through interactions with light or any other particles, a corollary is that a patch corresponding to the observable universe today was increasingly disconnected in the past. 
In particular, at the cosmic microwave background (CMB) last scattering surface, the patch size was about 1000 times  larger than the horizon size; see inset of figure~2.  Nevertheless, the CMB measurements show that the density and temperature were nearly uniform across the entire patch.
Explaining the uniformity of the CMB over lengths scales greater than the horizon size at that last scattering surface and at all previous times constitutes the {\it horizon problem}.  

The CMB measurements also reveal a spectrum of small amplitude density fluctuations that are nearly scale-invariant.  This inhomogeneity includes hot and cold spots that subtend patches substantially larger than horizon size at the time of last scattering.  Finding a physical mechanism for generating the scale-invariant spectrum, especially the part of the spectrum on wavelengths larger than the horizon size, constitutes the {\it inhomogeneity problem}.

The CMB temperature power spectrum also indicates that the  cosmic curvature factor, $\Omega_c(t_0)$ is less than 0.3\% today.  This is puzzling since the cosmic curvature factor $\Omega_c \propto (a^{2 \epsilon}/a^2)$ (or, equivalently, the ratio of the horizon size to the patch size) grows exponentially between  the Planck time and today (during which $a(t)$ grows by a factor of $10^{33}$).  That is, the initial spatial curvature as the universe emerges from the big bang is hugely amplified during this period.  The only way to explain the observed bound $\Omega_c(t_0) < 0.3\%$ today within the standard big bang paradigm is if the initial curvature was zero to many decimal places. Explaining this unexpectedly tiny initial value constitutes the {\it flatness problem}.

Long wavelength gradient energy in all degrees of freedom scales as $1/a^2$, like spatial curvature, in the linear regime, so the same considerations apply to explaining why the early universe was so uniform.  Resolving the horizon problem, {\it i.e.}, causal connectedness,  is not sufficient to explain uniformity; there must also be a smoothing mechanism that suppresses gradient energy.  Identifying this mechanism constitutes the {\it smoothness problem}.  Mechanisms that resolve the smoothness problem generically flatten the universe as well, so the two problems can be viewed as one; but this is not the current convention.

By contrast, the cosmic anisotropy factor  $\Omega_a \propto (a^{2 \epsilon}/a^6)$ decreases during a matter ($\epsilon =3/2$) or radiation ($\epsilon=2$) dominated expanding phase, and so there is no amplification problem.

In sum, the wedge diagram for the standard big bang model  illustrates all of the fundamental conundra that plague standard big bang cosmology: the cosmic singularity,  horizon, inhomogeneity, flatness, and smoothness problems.  
These all arise from the fact that classical spacetime as described by Einstein gravity `ends' at the vertex (the big bang) and  that the ratio of the horizon size to the patch size shrinks when extrapolating back in time.

\section{Non-singular bouncing cosmology}

Next we consider the wedge diagram for a generic classical (non-singular) bouncing cosmology.
By adding a contracting phase and bounce, the the causal structure is fundamentally changed. 
Now, as the diagram illustrates, there is a period before the bounce (far left) when the patch size is much smaller than the horizon size, which means the entire observable universe was causally connected in the past.  (Note that the horizon size (the shaded region) is defined to be $H^{-1}$, as before; however, in bouncing cosmology, the particle horizon, the set of all spacetime causally connected at {\it any} time in the past, is generally much larger.)

The diagram is different from simply gluing together a big bang and its time reverse in three important ways (see Figure~3):
\begin{figure}[tbp]
  \centering
\includegraphics[width=.85\linewidth]{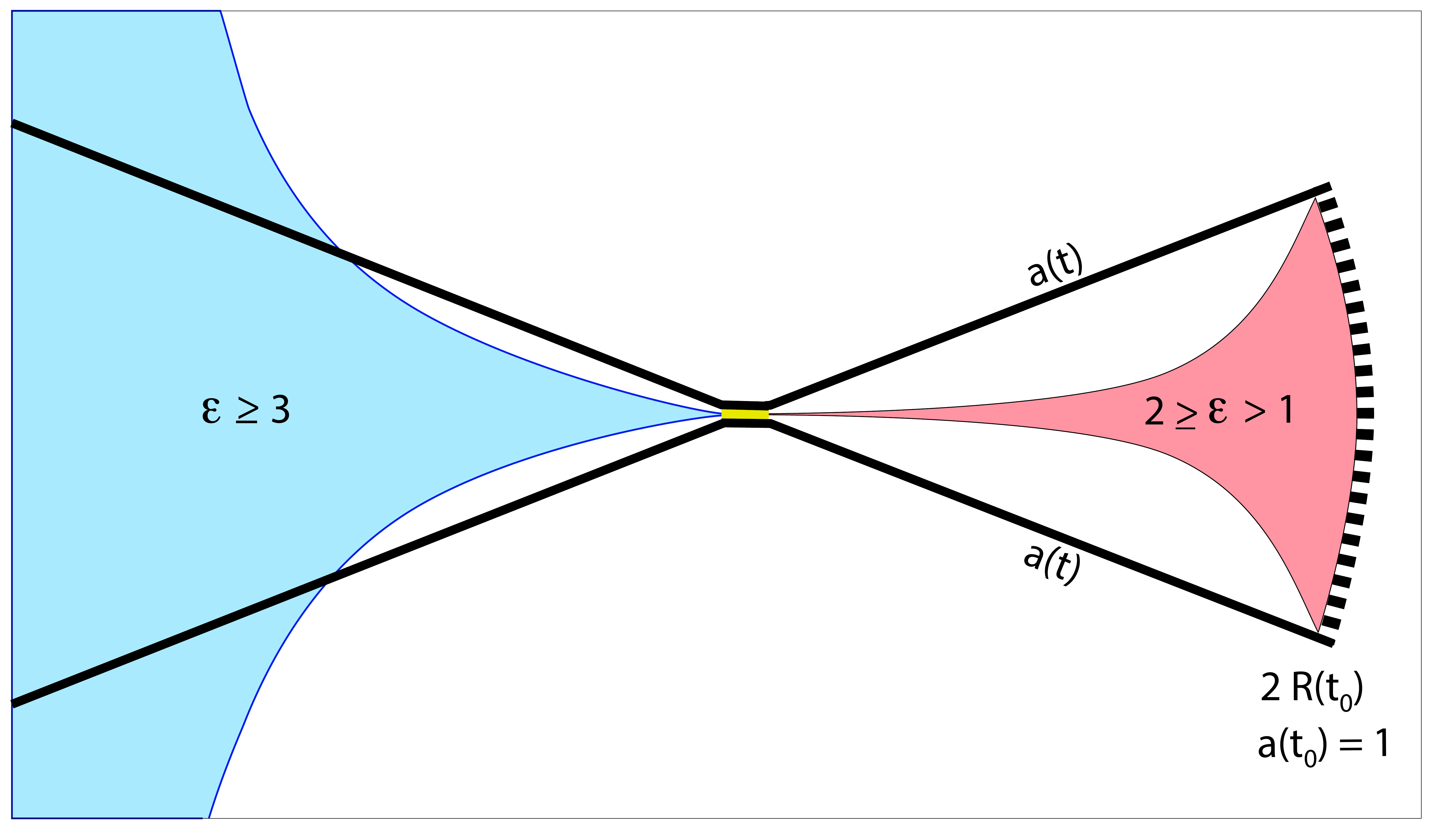}
\caption{Wedge diagram for non-singular bouncing cosmology.  A period of contraction (lhs) is followed by a bounce (small shaded connector, middle) and the current period of expansion (rhs).  The figure illustrates that the patch size (thick solid lines) is  larger than  the horizon size (shaded region) during the expanding phase, but was smaller than the horizon size during most of the contracting phase (lhs of diagram). Note that the evolution of the horizon size ($\propto a^{\epsilon}$)  is different before and after the bounce (shaded regions) because  $\epsilon \ge 3$ leading up to the bounce and, after conversion of energy to ordinary matter and radiation,  $2 \ge \epsilon $ after the bounce.  
} 
\label{fig:3}
\end{figure} 
\begin{itemize}
\renewcommand{\labelitemi}{--}
\item the equation of state $\epsilon$ is generically greater in the period before the bounce than in the period after the bounce;
\item  the wedge diagram need not terminate on the left-hand side.  In the case of a  one-time bounce model considered in this paper, the wedge might extend arbitrarily far and grow infinitely wide as $a \rightarrow -\infty$;\footnote{
In a cyclic model, the extrapolation would include regularly repeating periods of expansion, contraction and bounce.}
\item  the cosmic singularity is replaced by a bounce in which the scale factor shrinks to a {\it finite} critical size, well above the Planck length, and rebounds.
\end{itemize}

As the scale factor $a(t)$ decreases during a period of contraction, the energy component with the largest value of $\epsilon$ or, equivalently, the largest pressure comes to dominate the Friedmann equation Eq.~\eqref{F1} compared to forms of stress-energy with lower pressure because the energy densities scale as $1/a^{2 \epsilon}$ and $a$ is shrinking.  In the case of scalar fields, for example, because the blueshift effect during a period of contraction causes the kinetic energies of particles and fields to grow to the point where masses and potentials can become negligible.  
In this  limit,  the kinetic energy scales as $1/a^6$ and $\epsilon \rightarrow 3$.   

As a result, the cosmic anisotropy factor $\Omega_a \propto a^{2 \epsilon}/a^6$ in  Eq.~(\ref{anis}) becomes locked at a fixed value as the bounce approaches, naturally preventing the anisotropy factor from growing large enough to trigger chaotic mixmaster behavior \cite{Belinsky:1970ew}  that would otherwise disrupt the smooth contraction.  Models that reach the condition $\epsilon =3$ are, therefore, sufficient to obtain a smooth and isotropic universe before the bounce, and also turns out to be sufficient for generating a nearly scale-invariant spectrum of super-horizon perturbations \cite{Levy:2015awa}.  As a microphysical model, these models typically require fewer parameters and somewhat less tuning than cases with $\epsilon>3$, although the latter were studied first historically and may have certain advantages.\footnote{For example, mixmaster  behavior is even more strongly  suppressed for a scalar field with $\epsilon >3$.   In this case, $\Omega_a \propto a^{2 \epsilon}/a^6$  approaches zero, rather than a constant, as the bounce approaches.  The condition $\epsilon >3$ can occur, for example, for a scalar field with negative potential energy density during the bounce phase.}

According to the bouncing scenario, at some point during or shortly after the bounce, the kinetic energy stored in scalar  fields is converted to the matter and radiation we observe, with $\epsilon \leq 2$.   The irreversible reheating process accounts for the asymmetry in $\epsilon$ about the bounce point. 

Because the horizon size is proportional to $a^{\epsilon}$ and $\epsilon$ is larger before the bounce than after, the growth in the horizon size going away from the bounce is asymmetric, as represented figuratively in Figure~3, though the actual difference is exponentially greater quantitatively.  If, say, the bounce occurs at temperatures near $10^{15}$~GeV, then the scale factor must grow by a factor of nearly $e^{60}$ (or 60 ``$e$-folds'') before the horizon size grows to equal the patch size in the expanding phase, as shown on the right of the figure. On the other hand, extrapolating backwards in time from the bounce into the contracting phase, the horizon size catches up to the patch size after a change in scale factor by  $\approx 10^{15}$ (or 35 $e$-folds)  if $\epsilon =3$ and less if $\epsilon>3$.  Figuratively, this is represented in Figure~3 by having  the horizon size boundary be different on the left-hand and right-hand side of Figure~3.

Furthermore, as noted  above, the wedge diagram may continue arbitrarily far to the left such that, once the Hubble horizon catches up to the patch size going back in time, it continues to grow rapidly compared to the patch size forever into the past.  That means the patch that will eventually evolve into today's observable universe is not only causally connected at some point before the bounce  ({\it i.e.}, within a particle horizon), but it remains within the Hubble horizon and causally connected arbitrarily far into the past.  This strong causal connectedness enables smoothing processes to act over the entire region for an arbitrarily long period of time. 

Finally, a key ingredient in {\it non-singular} bouncing cosmology is a classically stable bounce that occurs when the energy density is well below the Planck density were quantum gravitational fluctuations are small.  As is well-known, a bounce of this type requires a modification of Einstein general relativity and/or a form of stress-energy  that violates the null energy condition ($\dot{H} \propto -(p+\rho) \le 0$).  Significant advances have been made recently in producing explicit examples of theories with stable, smooth, classical (non-singular) bounces from contraction to expansion.  We will comment on their status in Section~6 below.

\section{Generic features of classical (non-singular) bouncing cosmology}

In this section, we use the wedge diagram to show how the three features described in the previous section --  a contraction phase with $\epsilon \ge 3$; the continuation of the contraction phase arbitrarily far into the past; and the replacement of the big bang with a classical non-singular bounce -- combine to make a scenario that addresses many of the longstanding puzzles of cosmology.  We also assume that, during or shortly after the bounce, the energy density that dominated during the contracting phase is converted into ordinary matter and radiation.

For simplicity, we will only consider the case of a single (one-time only) bounce separating a semi-infinite period of contraction from the current period of expansion that began 13.8 Gyrs ago, although cyclically bouncing models are also possible in principle \cite{Steinhardt:2001st,Steinhardt:2004gk}.  


\begin{itemize}[leftmargin=*]
\renewcommand{\labelitemi}{\tiny$\blacksquare$}
\item {\bf geodesic completeness:}

As is apparent in Figure~3, every co-moving particle in the observable universe has a world-line that can be traced back through the bounce arbitrarily far back in time.  In the limit, the cosmology is past geodesically complete, unlike cosmologies based on the big bang.  

\item {\bf avoiding the cosmic singularity problem:}

By design, the energy density in {\it non-singular} bouncing cosmologies is at all stages far below the Planck density where quantum gravity effects are expected to be large.   This raises the thought-provoking and potentially empirically testable possibility that the large-scale properties of the observable universe are insensitive to details of quantum gravity.  Of course, understanding quantum gravity and unification remains an important goal for theoretical physics; but our point here is that quantum gravity may not play any direct role in shaping the large-scale structure of the observable universe.  A corollary is there is no quantum-to-classical transition to be explained. Evolution is dominantly classical at all stages.

\item {\bf eliminating chaotic mixmaster behavior:}

Chaotic mixmaster or BKL-like behavior can potentially destroy the smoothness of the universe during a period of contraction \cite{Belinsky:1970ew}. 
However, as noted above, mixmaster behavior is totally suppressed because $\epsilon \ge 3$ during the contraction phase \cite{Erickson:2003zm,Garfinkle:2008ei}.  

\item {\bf achieving causal connectedness:}

Considering now the complete bundle of world-lines associated with all co-moving particles in the observable universe, we can see from Figure~3 that they spend a semi-infinite period of contraction within a common, causally connected, cosmological horizon; then there is a finite interlude before and after the bounce when the horizon size becomes so  small compared to the patch  that most world-lines lie outside the horizon; finally, the horizon size grows fast enough to re-encompass them today.  For most of past cosmic history, though, the patch corresponding to the observable universe has been causally connected, a generic property of bouncing cosmology.

\item {\bf resolving the horizon problem:}

The horizon problem arises in standard big bang cosmology precisely because the patch corresponding to the observable universe was never causally connected in the past.  As  explained above and shown in Figure~3, this is not the case for bouncing cosmology. More specifically, in standard big bang cosmology,  the ratio between horizon size to the patch size (${\cal \eta} \equiv a^{\epsilon}/a$) decreases monotonically extrapolating back in time as $a$ decreases.  This trend automatically reverses, though, if the extrapolation is followed  through the bounce and into the contracting phase.  The horizon size grows arbitrarily larger than the patch size, so there is no causal impediment to having a smooth uniform universe. 

\item {\bf explaining the smoothness of the universe:}

Removing the causal impediment is necessary but not sufficient to explain why the energy density distribution was so smooth at the time of last scattering, early in the current expanding phase.  An explicit smoothing mechanism is needed.  In bouncing cosmologies, there is a generic smoothing mechanism associated with the natural behavior of  ${\cal \eta}^2 = a^{2\epsilon}/a^2$, which corresponds to the square of the ratio of the horizon size to the patch size, as pictured in Figure~3 moving left to right in the figure.  The ratio  ${\cal \eta}^2$ is key because it also determines the growth of gradient energy compared to the homogeneous contributions to the Friedmann equation in Eq.~\eqref{F1}.  In an expanding phase, the ratio increases, but, in a contracting phase, the ratio decreases. To explain the smoothness of the universe  after the bounce, it is only necessary that ${\cal \eta}^2$ 
decreases much more during the contracting phase leading up to the bounce than it increases during the expanding phase since.  This condition is naturally satisfied for two reasons. First, there is no bound to how much $a(t)$ can shrink during the contracting phase, but there is a bound to how much it could have increased between the bounce and today.  Second, the average value of $\epsilon$ during the contracting phase is greater than during the expanding phase.  Together, these two effects make the contraction phase an exponentially powerful smoothing mechanism.  

\item {\bf explaining the flatness of the universe:}

The same arguments apply to explaining the flatness of the universe, since the cosmic curvature factor $\Omega_c \propto {\cal \eta}^2$.   By the time the universe reaches the bounce, the factor becomes tiny and the spatial curvature is exponentially suppressed, as required to account for the flatness observed today.

\item {\bf generating super-horizon scale fluctuations:} 

Fluctuations of scalar fields on super-horizon scales (as perceived during the expanding phase) can be explained by quantum fluctuations that begin on sub-horizon scales evolving to subtend scales larger than the horizon by the time the universe reaches the bounce.  Note that the ratio of the horizon size to the patch size, ${\cal \eta}$, is also proportional to  the ratio of the horizon size to the wavelength of a quantum fluctuations, since the latter scales as $a(t)$.  The ratio shrinks exponentially as the universe approaches a bounce; hence, super-horizon modes with wavelengths much larger than the horizon size are naturally generated; see Figure~4. Since the generation of quantum perturbations and the contraction process are scale-free, the spectrum of super-horizon scale fluctuations is also scale-free generically \cite{Ijjas:2013sua}.  

The subtlety is converting the scalar field fluctuations into density (or curvature) fluctuations that explain the temperature variations in the CMB and the seeds for galaxy formation \cite{Creminelli:2004jg,Lehners:2007ac,Buchbinder:2007ad}. The conversion to density fluctuations with the precise tilt and other properties we observe is model-dependent.  Although some of the first examples of bouncing models in the literature produced significant non-gaussianity \cite{Buchbinder:2007at}, it was later discovered that this is not generic. The significant non-gaussianity turned out to occur in cases in which the background field trajectory is unstable.  Today simpler models with stable background field trajectories are known that  
 produce spectra with negligible non-gaussianity.  The ${\cal O}(1)$ parameters can be straightforwardly selected to produce spectral parameters consistent with current observations \cite{Ijjas:2014fja}. 
\begin{figure}[tbp]
  \centering
\includegraphics[width=.85\linewidth]{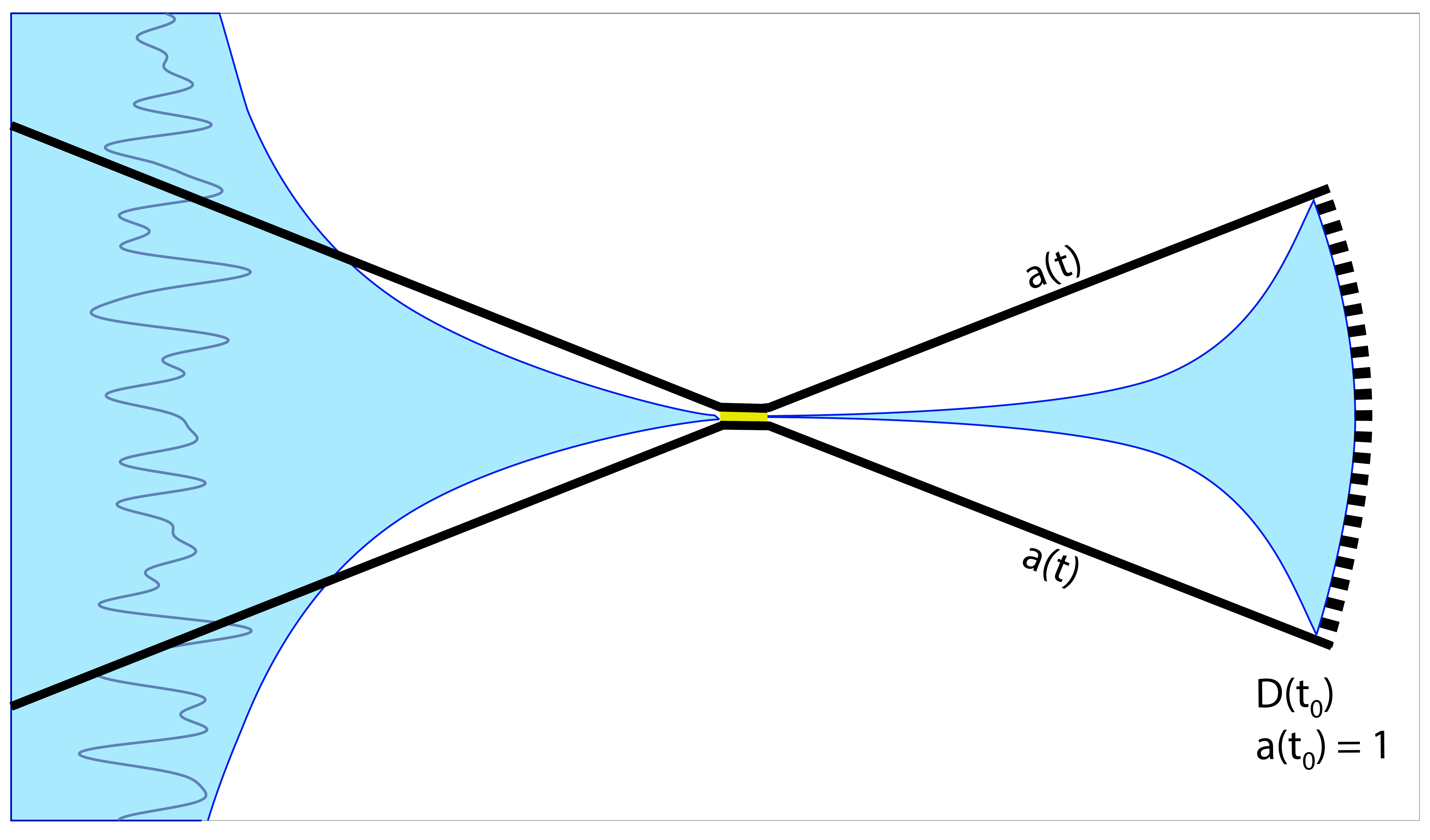}
\caption{Wedge diagram for non-singular bounce cosmology indicating the quantum fluctuations of a scalar field (wiggly curve) at a time during the contracting phase when the horizon size (shaded) is larger than the patch size (arc length across the thick solid lines).   Extrapolating through the bounce, these are ultimately the source of superhorizon density fluctuations (see text). } 
\label{fig:4}
\end{figure} 

\item {\bf suppressing the amplitude of (primary) tensor fluctuations:}

The quantum fluctuations responsible for the temperature variations observed in the CMB are produced in bouncing models with roughly 60 $e$-folds of contraction in $a(t)$ remaining {\it before} the bounce when the horizon size ($\propto a^{\epsilon}$) is exponentially large compared to the Planck scale.  According to the Friedmann equation, this means  the total energy density is exponentially small compared to the Planck density.  This corresponds to an arc far to the left in Figure~4.  If the total energy density or Hubble parameter is small, the amplitude of the associated tensor modes -- known as the primary tensor fluctuations on CMB scales -- is proportionately small, which explains why these models do not predict observable B-modes \cite{Boyle:2003km}.  (Late-time secondary tensor modes created when density fluctuations re-enter the horizon in the matter-dominated universe occur independent of the mechanism that generated the density fluctuations in the first place; so they also occur generically in bouncing models.  They correspond to an effective tensor-to-scalar ratio of $r \lesssim 10^{-6}$ which produces B-modes that are too small to detect with current techniques \cite{Baumann:2007zm}.)

\item {\bf avoiding quantum runaway (a.k.a. the multiverse problem):}

The smoothing phase in bouncing models is the phase of contraction. The wedge diagram in Figure~3 gives the impression that the evolution is perfectly uniform, but this is only an approximation.
Quantum fluctuations on super-horizon scales change the local rate of contraction before the bounce, and, hence, can advance or retard when smoothing and contraction ends; see Figure~4.  

Quantum runaway is a phenomenon that {\it generically} occurs if gravitational effects somehow amplify the volume of rare large fluctuations and keep such regions in the smoothing phase long enough to produce yet more rare fluctuations on top of that.  Inflation is a well-known example \cite{Steinhardt:1982kg,Vilenkin:1983xq,Linde:1986fd,Guth:2007ng}. 

Quantum runaway does {\it not} occur in bouncing models because the analogous rare fluctuations that keep regions in the smoothing phase create patches that are contracting, leaving less volume and insufficient time for additional rare fluctuations.

\item {\bf explaining the small entropy at the beginning of the expanding phase:}

Bouncing cosmology satisfies the second law of thermodynamics.  The entropy density of the universe today  is exponentially smaller than the maximum entropy possible, which means it was only smaller still at the beginning of the expansion phase.  The wedge diagram in Figure~3 suggests a possible reason why that might be so.  Namely, the patch corresponding to our observable universe today was only an infinitesimal fraction of the horizon size long before the bounce.  That means only the limited entropy in the pre-bounce phase that is contained within narrow wedge contributes to what is in the observable universe at the beginning of the expanding phase.   (Similarly, in cyclic versions of bouncing cosmology, the region that evolves into the observable universe a cycle from now occupies today a volume that is only a few meters across \cite{Steinhardt:2001st}.  As a result, only the limited entropy that is contained within that few-meter volume today  contributes to the entropy an observer would see a cycle from now.  The rest lies in regions of space outside the observer's horizon.)
\end{itemize}

\section{Final remarks}

The tantalizing possibility of having a cosmological scenario with all the properties above
is strong motivation for exploring classical (non-singular) bouncing cosmologies.  Perhaps the most significant challenge in constructing examples is rigorously establishing mechanisms for the bounce.   A fully-computable bounce model is indispensable to show that smooth and flat physical patches with scale-invariant, super-horizon density fluctuations produced in the contracting phase {\it generically} transit to expansion.
More precisely, a smooth classical  bounce means changing the  monotonic behavior of the scale factor $a(t)$ from decreasing (contraction) to increasing (expansion) at sub-planckian energies  well before $a(t)$ shrinks to zero.

While the contracting phase is minimalist in the sense that it can be realized by assuming conventional Einstein gravity and simple forms of stress-energy, such as a canonical scalar-field  with large kinetic energy density, a smooth classical bounce requires  a stable violation of the null energy condition and/or modifications of Einstein gravity.  Such modifications could naturally occur at  energy scales reached near the bounce, which are well below the Planck scale where quantum gravity corrections become important but higher than the energy-scales reached during the contracting phase. 
However, in constructing smooth bounce models, one must be aware of potential pitfalls, such as wrong-sign kinetic or gradient terms, and singularities.\footnote{A wrong-sign kinetic term in the action for physical (gauge-invariant) quantities introduces a ghost; this is a {\it quantum} instability. Wrong-sign gradient terms, on the other hand, lead to an imaginary sound speed. This problem has often been misleadingly dubbed a `gradient instability' in the literature.  This term suggests that the problem can be resolved if the instability lasts for a short time and only affects long-wavelength modes based on the notion that the `instability' evolves too slowly for those  modes to adversely affect the outcome. In reality, though, what happens is that the dynamical equation for the corresponding physical quantity fundamentally changes form from strongly hyperbolic to elliptic.   As a consequence, the dynamics becomes acausal in the sense that it depends on future boundary conditions. A corollary is that the pathology is immediate if a wrong-sign gradient term is ever encountered for super-planckian wavelengths; a well-behaved bounce must avoid wrong-sign gradient terms on long wavelengths altogether.} 
At present, several approaches are being investigated. (For some recent examples see, {\it e.g.}, \cite{Cai:2016thi,Creminelli:2016zwa,Kolevatov:2017voe,Farnsworth:2017wzr,Graham:2017hfr}.)  The most advanced, based on braided scalar-tensor modifications of gravity, is highly promising because it has been shown to evade all the known pitfalls at the perturbative level; see \cite{Ijjas:2016vtq,Ijjas:2017pei}. The next step is to establish stability and smoothness at the non-linear level and search for any observational imprints \cite{Ijjas:to-appear}.

Another promising possibility for bouncing cosmologies is embedding them in cyclic theories of the universe in which bounces recur at regular intervals.  These theories do not just describe the early evolution of the universe, but its entire history.  Consequently, recent stages, such as dark matter and dark energy domination, are naturally closely tied to bounces both past and future, imposing novel qualitative and quantitive constraints that can make bouncing cosmology more powerfully predictive.  For example, one immediate prediction of cyclic theories is that the current dark energy dominating phase must be metastable or slowly decaying, ultimately transitioning to a state of lower energy density that will initiate a period of contraction.  Cycling may also explain the magnitude of the dark energy density and other fundamental parameters \cite{Steinhardt:2006bf,Lehners:2010ug}.

\section*{Acknowledgements}

We thank D. Hogg and M. Rozenblit for useful comments; and PJS thanks the NYU Center for Cosmology and Particle Physics for generous hospitality and support during his leave when this work was completed.
This research was supported by the Simons Foundation Program `Cosmological Bounces and Bouncing Cosmologies,'  as part of the `Origins of the Universe' Initiative. PJS is supported by the DOE grant number  DEFG02-91ER40671.

\appendix
\section*{Appendix: Big bang inflationary cosmology}
\addcontentsline{toc}{section}{Appendix: Big bang inflationary cosmology}

Inflation \cite{Guth:1980zm,Linde:1981mu,Albrecht:1982wi} has been proposed as an enhancement of standard big bang cosmology that only alters cosmic evolution during the first $10^{-30}$~sec or so after the bang.   Consequently, the wedge diagram would appear the same as in Figure~2 given the coarse-grained resolution of $a(t)$ shown in the figure.  The difference would only become apparent by exponentially magnifying the region near the vertex by a factor of 60 $e$-folds ($\approx 10^{26}$) or so.

This is pictured in Figure~5, where the inflationary phase is illustrated as part of a blow-up of the extreme corner of the wedge diagram.  (The inset should be regarded as a qualitative sketch; it is not practical to show the actual exponential range of scales involved.)
The narrow wedge represents the patch corresponding to  the observable universe today extrapolated back to these early moments.  The wider wedge describes the evolution of a patch that was equal to the horizon at the beginning of inflation,  represented by the left hand arc bounding the shaded region marked $\epsilon <1$. The fact that it is wider means that the observable universe was smaller than the horizon size when inflation began and that the initial horizon-sized region has inflated to an enormous size that includes our observable universe today.
\begin{figure}[tb]
  \centering
\includegraphics[width=.7\linewidth]{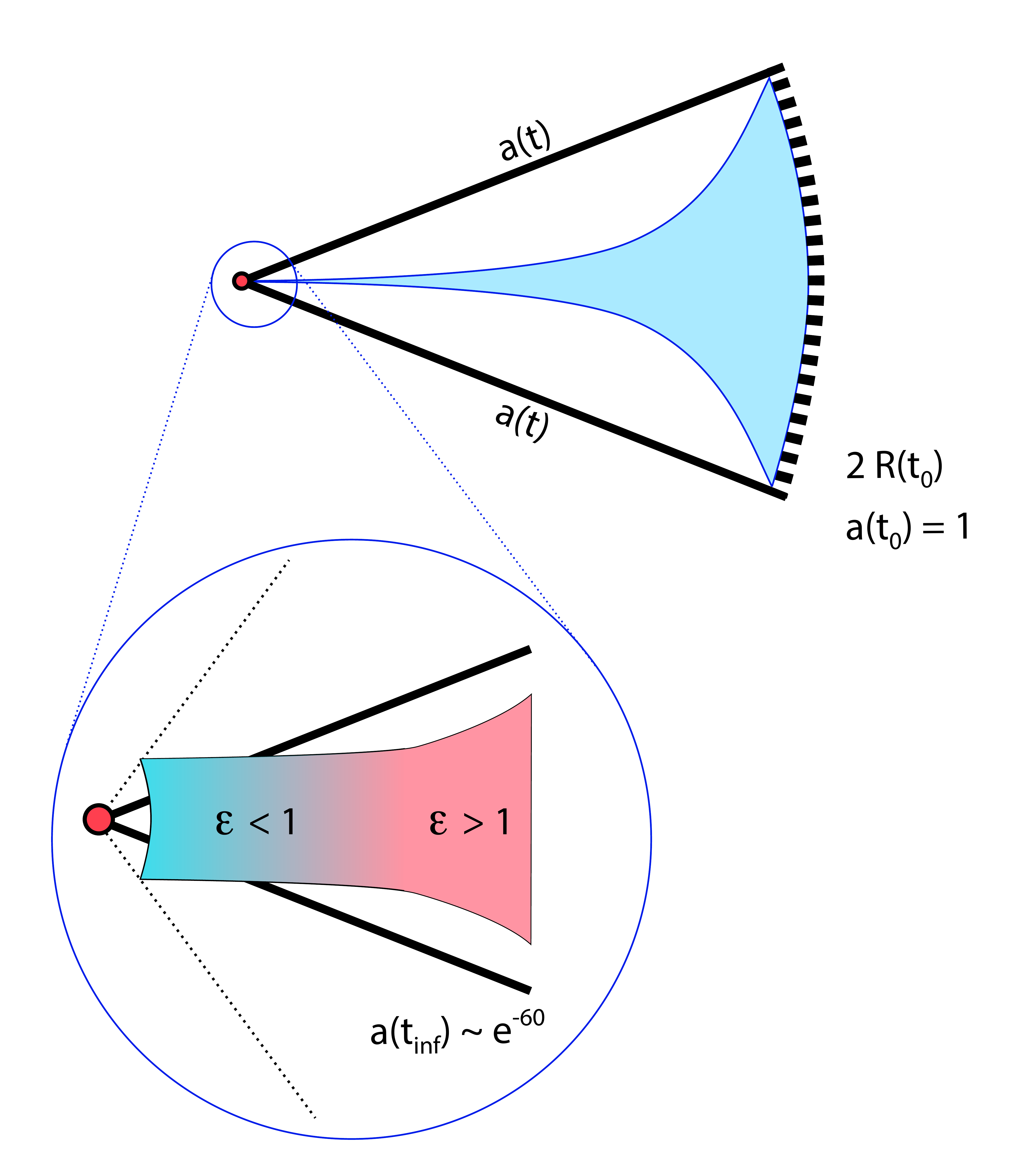}
\caption{Wedge diagram of big bang inflationary model.  The main figure shows the big bang wedge as in  Figure~2; the inset magnifies the region very close to the big bang that includes the inflationary epoch ($\epsilon <1$) and the reheating epoch ($\epsilon >1$) that follows.  Assuming 60 or more $e$-folds of inflation, the horizon size at the beginning of inflation (arc on lhs of shaded region) corresponds to a wedge (dotted lines) that is wider than the wedge (thick lines) that evolves to the size of the observable universe today.  The characteristic feature of inflation is that, evolving towards the rhs, the inflating patch size grows to be larger than the horizon size.  Note that inflation is preceded by a singularity and there is a  gap between the big bang and the beginning of inflation to be explained. } 
\label{fig:4}
\end{figure}   

Note that both wedges meet at the vertex, as in the standard big bang diagram. This demonstrates the fact that {\it inflation does not avoid the cosmic singularity problem}.  It simply inserts a special period of evolution (the region marked $\epsilon<1$) between the singularity and the periods of radiation- and dust-domination (the region marked $\epsilon>1$).  

Without inflation, $\eta = a^{\epsilon}/{a}$ grows with $a$ because $\epsilon >1$.  That is, the horizon size shinks to zero faster than the patch size (the narrow wedge, as occurs on the right side of the figure where $\epsilon >1$).  Inflation or accelerated expansion occurs for forms of stress-energy for which $\epsilon <1$.  In that case, the horizon size shrinks more slowly than the patch size (the narrow wedge) so that eventually, the patch size becomes smaller than the horizon size. How much smaller it gets depends on the duration of inflation, which in this figure depends on how far the beginning of inflation (the left hand arc bounding the region marked $\epsilon <1$) is pushed towards the singularity. 

When the patch size (the narrow wedge) becomes smaller than the horizon size, our observable universe is causally connected, which is how inflation attempts to get around the causal connectedness issue). Furthermore, quantum fluctuations generated on the horizon scale at that time extend over length scales wider than the narrow wedge; these expand along with the wedge going forward in time and become super-horizon scale fluctuations.  

The cosmic curvature factor, $\Omega_c \propto a^{2 \epsilon}/a^2$, shrinks exponentially during the inflationary phase where $\epsilon<1$, and  space is thereby flattened.   After reheating, the cosmic curvature factor  grows  again during the radiation- and dust-dominated phases that follow because $\epsilon>1$.  The flattening during inflation must last long enough (60 $e$-folds of expansion) to overcome the growth in the curvature factor that follows. This is the inflationary approach for resolving the flatness problem.

In addition to showing how inflation works, the wedge diagram can be used to illustrate some of the open issues and challenges.  As noted in the main text, the natural progression in an expanding (or contracting) universe is for $\epsilon$ to increase as $a(t)$ approaches zero because the term with the highest value of $\epsilon$ tends to dominate as $a \rightarrow 0$ in the Friedmann equation, Eq.~(\ref{F1}).  This is the sequence that occurs in $\Lambda$CDM (extrapolating back in time towards smaller $a$, first $\Lambda$, $\epsilon =0$ dominates; then matter, $\epsilon=3/2$;  and then radiation, $\epsilon =2$); the same applies in the bouncing model.  Inflation disrupts this order by inserting a period with $\epsilon <1$ before the radiation- and dust-dominated phases with $\epsilon >1$.  Disrupting this order is not generic in standard big bang cosmology; it requires fine-tuning of initial conditions.    

For example, in models of inflation driven by a scalar field rolling down a potential, inflation only occurs if the scalar field  potential energy density is large ($\epsilon \approx 0$)  compared to its kinetic energy density ($\epsilon =3$).  Near the big bang, though, this simple $\epsilon$-analysis shows that the reverse is the generic case: kinetic energy density tends to dominate exponentially over potential energy.  To have a slowly-rolling scalar field requires exponential fine-tuning. 
 
Furthermore, note that the gap between the cosmic singularity and the beginning of inflation (the left hand arc bounding the region marked $\epsilon <1$).  This captures the fact that the inflationary phase is geodesically incomplete -- it must have a beginning that occurs at some point {\it strictly after} the big bang and there must be some period of non-inflationary evolution in between.

The wedge diagram can be used to illustrate this fact.  Recall that, {\it formally}, space disappears at the singularity; hence, the patch size (both narrow and wide wedges) and the horizon size must all reach zero at the same point in the diagram.  The only way this can happen geometrically in the diagram is if, at some point leading up to the singularity, the horizon size shrinks as fast or faster than  the patch size as $a \rightarrow 0$. The patch size shrinks in proportion to $a$. The horizon size must shrink as $a^p$ where $p\ge 1$.  This corresponds to $\epsilon \ge 1$, which, by definition, cannot be inflation.   Of course, if it is not inflation, then, just as in the standard big bang model, there is no reason to expect for the universe to evolve from the big bang to the relatively smooth conditions needed to begin inflation, which adds to the inflationary initial conditions problem.  

Summing up, neither the small $\epsilon$ ({\it e.g.}, small scalar field kinetic energy density) or smoothness essential for inflation to start are generic outcomes of a big bang universe.  Finding a resolution to this conundrum is a major open issue for inflationary cosmology.

Another well-known feature of inflationary expansion is the quantum runaway effect that leads to a multiverse.  Patches that undergo rare large quantum fluctuations significantly delaying the end of inflation grow exponentially large in volume, enough to create patches within that undergo rare fluctuations and keep inflation going yet longer and over yet larger volumes, etc.  Patches that end inflation have cosmological properties depending on the rare fluctuations that preceded them. This allows for all possible outcomes with no firm statistical rationale for preferring one outcome over another.   Resolving the quantum runaway problem is a challenge for big bang inflationary cosmology, though one not easily represented in a wedge diagram.

\bibliographystyle{apsrev}
\bibliography{bouncecase}

\end{document}